\documentstyle[preprint,prl,aps]{revtex}

\begin{document}
\draft

\title{Response Functions Improving Performance in Analog Attractor Neural
Networks}

\author{Nicolas Brunel\cite{nb}}
\address{INFN, Dipartimento di Fisica, P.le Aldo Moro 2,
00185 Roma, Italy}

\author{Riccardo Zecchina\cite{rz}}
\address{Dip. di Fisica Teorica e INFN, Universit\`a di
Torino, Via P.Giuria 1, 10125 Torino, Italy}

\date{\today}
\maketitle
\begin{abstract}
In the context of attractor neural networks, we study how the equilibrium
analog neural activities, reached by the network dynamics during memory
retrieval, may improve storage performance by reducing the interferences
between the recalled pattern and the other stored ones. We determine a simple
dynamics that stabilizes network states which are highly correlated with
the retrieved pattern, for a number of stored memories that does not exceed
$\alpha_{\star} N$, where $\alpha_{\star}\in[0,0.41]$ depends on the global
activity level in the network and $N$ is the number of neurons.

\end{abstract}

\narrowtext

Attractor neural networks (ANN's) have been the subject of intensive study
among physicists since the original paper of Hopfield\cite{ho,amit}. The
analogy between the thermodynamics of ANN's and of spin glasses has been used
to interpret the associative processes taking place in neural networks in terms
of collective nonergodic phenomena. The identification of attractors with the
internal representation of the memorized patterns, though still an object of
open discussion, has received some basic experimental
confirmations\cite{miyashita} and represents one of the basic issues for the
biologically motivated models currently under study.
Several authors have studied ANN's composed of analog neurons instead of the
discrete spinlike neurons of the original model \cite{ho2,at,shiino,kuhn} and
have shown that such more realistic networks may perform as associative
memories.
In the present paper we will be concerned with the following issue: assuming
the interaction (synaptic) matrix to be the simple Hebb-Hopfield correlation
matrix, we discuss how the storage performance of an ANN may depend on the
equilibrium analog neural activities reached by the dynamics during memory
retrieval.

In both discrete and analog Hopfield-like attractor neural networks, the
phase transition of the system from associative memory to spinglass
is due to temporal correlations arising from the static noise produced by the
interference between the retrieved pattern and the other stored memories.
The introduction of a suitable cost function in the space of neural activities
allows us to study how such a static noise may be reduced and to derive a class
of simple response functions for which the dynamics stabilizes the
``ground-state'' neural activities, i.e., the ones that minimize the cost
function, for a macroscopic number of patterns.

In what follows, we first give some basic definitions and successively do
the following.

{\bf (i)} We study the ground states of a cost function $E$ defined in the
phase space {\large $\varepsilon$} of the neural activities and proportional to
the sum of the squared overlaps of the network state with the stored memories
except the retrieved one.

{\bf (ii)} We derive the associated gradient flow in {\large $\varepsilon$} and
show that it converges to the ground state of $E$.

{\bf (iii)} We show that when the minimum of the cost function is zero there is
a linear relation between the afferent current and the activity at each site of
the network.

{\bf (iv)} We determine a simple dynamics (which turns out to be characterized
by a nonmonotonic response function) stabilizing the ground state of the
system when the number of stored memories is smaller than $\alpha_{\star} N$,
where $\alpha_{\star}\in[0,0.41]$ depends on the global activity level.

The neural network model is assumed to be fully connected and composed of $N$
neurons whose activities $\{V_i\}_{i=1,N}$ belong to the interval $[-1,1]$.
The global activity of the network is defined by
\begin{equation}
\label{defact}
\gamma=\frac{1}{N}\sum_i \epsilon_i \; \;\; ,
\end{equation}
where $\epsilon_i=|V_i|$ and  we denote by {\large$\varepsilon$}$=[0,1]^N$ the
space of all $\epsilon_i$.
A macroscopic set of $P=\alpha N$ random binary patterns
$\Xi\equiv {\bf \{} \{\xi_i^{\mu}=\pm 1 \}_{i=1,N} \, , \mu=1,P {\bf \}}$,
characterized by the probability distribution
$P(\xi_i^{\mu})={1\over2}\delta(\xi_i^\mu-1)+{1\over2}\delta(\xi_i^\mu+1)$,
is stored in the network by means of the Hebb-Hopfield learning rule
$ J_{ij}=\frac{1}{N}\sum_{\mu=1}^{P}\xi_i^{\mu}\xi_j^{\mu}$ for
$i\neq j$ and $J_{ii}=0$.

At site $i$, the afferent current (or local field) $h_i$ is given by the
weighted
sum of the activities of the other neurons $ h_i=\sum_{j}J_{ij}V_j$. We
consider a continuous dynamics for the depolarization $I_i$ at each site $i$
[$\tau \dot{I}_i(t)=-I_i(t)+h_i(t)$],
in which the activity of neuron $i$ at time $t$ is given by $V_i(t)=f(I_i(t))$,
where $f$ is the neuronal response function.
No assumptions are made on $f$, except that it is such as to align the
neural activity with its afferent current [$x f(x) \geq 0$ for all $x$].

We consider the case in which one of the stored patterns, ${\mu=1}$ for
example,
is presented to the network via an external current which forces the initial
configuration of the network, thus we take $V_i=\epsilon_i\xi_i^1$.
The current arriving at neuron $i$ becomes
\begin{equation}
\label{field}
h_i=\gamma\xi_i^1+\frac{1}{N}\sum_{\mu\neq 1}\xi_i^{\mu}
\sum_{j\neq i}\xi_j^{\mu}\xi_j^1\epsilon_j
\end{equation}
which, in terms of the overlaps of the network state with the stored patterns
$m_{\mu} \equiv \frac{1}{N}\sum_{j}  \xi_j^{\mu} \xi_j^1 \epsilon_j$,
reads
\begin{equation}
\label{fieldover}
h_i=m_1\xi_i^1 +\left(\sum_{\mu>1}\xi_i^{\mu}m_{\mu}-\alpha
\epsilon_i\xi_i^1\right) \; \; \;.
\end{equation}
Notice that the global activity is equal to the overlap of the network
configuration with the retrieved pattern ($m_1=\gamma$).

The first term in the right-hand side (rhs) of Eq. (\ref{fieldover})
is the signal part, whereas the second term, when the $\epsilon_i$ are fixed
and for $N$ large, is a Gaussian random variable with zero mean and variance
$\sigma^2=\sum_{\mu>1} m_{\mu}^2$ and represents the static noise part, or
cross-talk, due to the interference between the stored patterns and the
recalled one.
In the following we will be interested in minimizing the overlaps of the
network state with the stored memories $\mu\neq 1$ which are not recalled.
If one succeeds in finding such states ($m_{\mu}=0$ for all $\mu\neq
1$), then the second term in the rhs of Eq. (\ref{fieldover}) reduces to
$-\alpha \epsilon_i\xi_i^1$, and the interference effect vanishes.
We will see that this is indeed the case in a finite range of the
parameter $\alpha$ and that it is also possible to derive a class of effective
response functions realizing such a a minimization of the interference and thus
leading to an improvement of the network storage capacity.

{\bf (i)} In order to study how the interference may be reduced, we
define a cost function  $E(\Xi,\vec{\epsilon})$ depending on the neural
activities and the set of  stored patterns, proportional to $\sigma^2$, i.e. to
the sum of the squared overlaps of the network state with all the stored
memories  except the retrieved one
\begin{equation}
\label{defe}  E(\Xi,\vec{\epsilon})=\frac{1}{N}\sum_{\mu\neq
1}\left(\sum_j  \xi_j^{\mu}\xi_j^1\epsilon_j\right)^2
\end{equation}
and we study its ground states.
If no constraints are present on $\vec{\epsilon}$,  the minimum is always $E=0$
for $\vec{\epsilon}={\bf 0}$; obviously, in this case no information is
obtained  when
one presents the pattern and therefore we impose the  constraint $\gamma=K$ on
the average activity, where $K\in[0,1]$.
The geometrical picture of the problem is indeed very simple. In the space
{\large $\varepsilon$} we have to find the vectors $\vec{\epsilon}$ as
orthogonal as possible to the $P-1$ vectors
$\vec{\eta}^{\mu}=\{\eta_i^{\mu}=\xi_i^1 \xi_i^{\mu}\}$, with the constraint
$\gamma=K$. If there exists (with probability 1) at least one
$\vec{\epsilon}$ orthogonal to the $P-1$ vectors satisfying the constraint,
then we have $E=0$. The subspace corresponding to the condition $E=0$ is
connected since it is the intersection of $P-1$ hyperplanes $m_\mu=0$ and one
hyperplane $\gamma=K$.

In order to determine the typical free energy, we compute $\langle\ln
Z\rangle_{\Xi}$, where $\langle \ldots \rangle_{\Xi}$ stands for the average
over the quenched random variables ${\Xi}$ and $Z$ is the partition function at
temperature $T=1/\beta$ given by $Z= {\rm Tr}_{\vec{\epsilon}}\{\exp[-\beta
E(\Xi,\vec{\epsilon})]\delta \left( \gamma(\vec{\epsilon})-K\right)\}$, using
the standard replica method. Starting from the typical partition function of
$n$ replicas of the system $\langle Z^n\rangle_{\Xi}$, for $n$ integer, we
perform an analytic continuation for non-integer values of $n$, thus obtaining
$\langle\ln Z\rangle_{\Xi} = \lim_{n\rightarrow 0} \frac{\langle
Z^n\rangle_{\Xi} -1}{n}$.

Each replica $a$ ($a=1,\ldots,n$) is characterized by its neural activities
$\vec{\epsilon}^a$ and by the order parameter
$Q^a=\frac{1}{N}\sum_i(\epsilon_i^a)^2$, whereas the overlap between neural
activities in two different replicas defines the other order parameters (for
$a<b$) $q^{ab}=\frac{1}{N}\sum_i\epsilon_i^a\epsilon_i^b$. We
indicate with $R^a$ and $r^{ab}$ the conjugate parameters of $Q^a$ and
$q^{ab}$,
respectively. The typical free energy $F$ per site is then given, in the
thermodynamical limit, by $F(\beta)=-G(\beta)/\beta$
where $G(\beta)=\lim_{N\rightarrow\infty}<\ln
Z(\beta)>_{\Xi}/N$. The free energy at zero temperature, $F_0$, gives the
ground state of the system. $G$ can be calculated using a saddle--point method
that, once a replica symmetric (RS) ansatz $q^{ab}=q$, $r^{ab}=r$ (for all
$a<b$), and $Q^a=Q$, $R^a=R$ (for all $a$) has been done, leads to

\begin{eqnarray}
G & = & \min_{\cal M}\left[\frac{1}{2}(rq+RQ+Ku)-
\frac{\alpha}{2}\frac{\beta q}{1+\beta(Q-q)}\right.\nonumber\\
 & & \left. -\frac{\alpha}{2}\ln[1+\beta(Q-q)]+\int D\zeta \ln
\Gamma_{u,r,R}(\zeta)\right] \; \; ,
\end{eqnarray}
where ${\cal M}=\{ q, r, Q, R, u \}$,
\begin{equation}
%\label{defGamma}
\Gamma_{u,r,R}(\zeta)=\int_0^1
dz\exp\left[-\frac{R+r}{2}z^2-\left(\frac{u}{2}+\sqrt{r}\zeta \right) z
\right] \; \; ,
\end{equation}
and $D\zeta$ is the Gaussian measure
$d\zeta g(\zeta)$ with $g(\zeta)=\exp(-\zeta^2/2)/\sqrt{2\pi}$.
Depending on the storage level $\alpha$ one obtains the following results.

(a) For $\alpha<\alpha_0(K)$ we have $F_0=0$.
The parameter $q$, which is the typical overlap
between the activity configurations in two replicas such that the free energy
vanishes, increases from $q=K^2$ at $\alpha=0$ to $q=Q$ at $\alpha= \alpha_0$.
At $\alpha=\alpha_0$, the space of neural activities such that $F_0=0$ shrinks
to zero, and $F_0$ becomes positive.

(b) For $\alpha>\alpha_0(K)$, we have $q=Q$ and thus, in the limit
$\beta\rightarrow\infty$, we introduce  the new scaled variables
$r=\rho\beta^2$, $R+r=\sigma\beta$, $u=2\omega\beta$, and
$x=\beta(Q-q)$.
The order parameters are given by the following saddle-point equations:
\begin{equation}
\label{sp1}
Q=\int_{\zeta_0}^{\infty}D\zeta+\int_{\zeta_1}^{\zeta_0}D\zeta\left(
\frac{\zeta-\zeta_1}{\zeta_0-\zeta_1}\right)^2 \; \; \; ,
\end{equation}
\begin{equation}
\label{sp2}
K=\int_{\zeta_0}^{\infty}D\zeta+\int_{\zeta_1}^{\zeta_0}D\zeta\left(
\frac{\zeta-\zeta_1}{\zeta_0-\zeta_1}\right) \; \; \; ,
\end{equation}
\begin{equation}
\label{sp3}
x\sqrt{\rho}=\int_{\zeta_0}^{\infty} \zeta D\zeta +
\int_{\zeta_1}^{\zeta_0} \zeta D\zeta \left(
\frac{\zeta-\zeta_1}{\zeta_0-\zeta_1}\right) \; \; \; ,
\end{equation}
\begin{equation}
\label{sp4_5}
\rho=\frac{\alpha Q}{(1+x)^2} \; , \; \; \sigma=\frac{\alpha}{1+x} \; \; \; ,
\end{equation}
with $\zeta_0=\sigma/\sqrt{\rho}+\zeta_1,\;
\zeta_1=\omega/\sqrt{\rho}$.
$\alpha_0(K)$ (Fig. 1) is obtained in the limit $x\rightarrow\infty$
and the minimum $F_0$ is given by $F_0=\alpha Q (1+x)^{-2}$ where $Q$ and $x$
are fixed by their saddle--point values.

The condition of local stability of the RS solution with respect to small
fluctuations in replica space has been calculated. It turns out to be verified
at $T=0$ for all $\alpha$, which is not surprising since, as previously
noticed, the space of neural activities that minimize the cost function is
connected.

The calculation of the probability distribution of the activities
in the ground state can be done with similar techniques and
yields
\begin{equation}
%\label{Pgen}
{\cal P}(\epsilon)=\int_{-\infty}^{+\infty}
\frac{D\zeta}{\Gamma_{u,r,R}(\zeta)}
\exp\left[-\frac{R+r}{2}\epsilon^2-
\left(\frac{u}{2}+\sqrt{r}\zeta\right)\epsilon \right] \; \; \; ,
\end{equation}
where $u$, $r$, and $R$ take their saddle--point values.
Above the critical storage level $\alpha\geq\alpha_0$, the probability
distribution reads, for $\epsilon\in[0,1]$,
\begin{equation}
%\label{Palpha}
{\cal P}(\epsilon)=H(\zeta_0)\delta(\epsilon-1)+ \Delta_\zeta g(\zeta_1+
\epsilon \Delta_\zeta)
+H(-\zeta_1)\delta(\epsilon) \; \; \; ,
\end{equation}
where $H(\zeta) \equiv \int_{\zeta}^{\infty} Dz$,
$\Delta_\zeta=\zeta_0-\zeta_1$
and $\zeta_0$ and $\zeta_1$ are given as a function of $\alpha$ and $K$ by
Eqs. (\ref{sp1}),(\ref{sp4_5}). Notice the two $\delta$ functions in 0 and
1.

For brevity, we do not report here the results\cite{bz2} concerning the case of
discrete (three-states $\{\pm 1,0\}$) neurons; we just mention that replica
symmetry breaking is required.

{\bf (ii)} The next step regards the gradient flow associated with a smooth
version  of the energy function (\ref{defe}), implementing a soft quadratic
constraint  for the global activity.
The study of this gradient flow will allow us, on the one hand,
to find the relation between activities and afferent currents in the ground
state, and on the other, to check the outcome of the RS solution.
We emphasize that this gradient flow does not correspond to the
dynamics of the network -- this point will be considered later in (iii) and
(iv).

The new cost function $E_{\lambda}$ can be written
\begin{equation}
\label{elambda}
E_{\lambda}(\Xi,\vec{\epsilon})=\frac{1}{N}\sum_{\mu\neq 1}\left(\sum_j
\xi_j^{\mu}\xi_j^1\epsilon_j\right)^2+\frac{\lambda}{N}\left(\sum_i\epsilon_i
-KN\right)^2 \; \; \; ,
\end{equation}
where $\lambda\!>\!0$ is a Lagrange multiplier.
In the first term one recognizes the previous cost function (\ref{defe})
whereas the second term is introduced in order to favor configurations with
activity $K$. The ground state $F_\lambda$ of $E_{\lambda}(\Xi,\vec{\epsilon})$
can be calculated\cite{bz2} with the same methods as the ground state of
(\ref{defe}). For $\alpha<\alpha_0(K)$ we have $F_{\lambda}=0$ and $\gamma=K$,
while for $\alpha>\alpha_0(K)$ the ground state $F_{\lambda}$ becomes positive,
and we have $\gamma<K$.
By computing the gradient of $E_{\lambda}(\Xi,\vec{\epsilon})$, it is now
easy to derive the flow in the
{\large $\varepsilon$} space: in $R^N$ we have
\begin{equation}
\label{gradE}
\dot{\vec{\epsilon}} = -\frac{\tau'}{2} \nabla E_\lambda = \tau'
(A \vec{\epsilon} + \vec{b}) \; \; \;, \; \; \;\tau'\!>\!0 \; \;
\end{equation}
where $\{A_{ij}\!\!=\!\!-(\frac{1}{N} \sum_\mu \eta_i^\mu
\eta_j^\mu-\frac{1-\lambda}{N})\}$ and $\{b_i\!\!=\!\!\lambda K\},\;
i,j=1,N$.
Discretizing time, constraining the $\epsilon_i$ to stay in the $[0,1]$
interval, and choosing $\tau'=\frac{1}{\alpha}$, we arrive at the following
local equation
\begin{equation}
\label{simpleupdating}
\epsilon_i(t+1) = \phi\left(\frac{1}{\alpha}\left\{
\lambda [K-\gamma(t)]+\gamma(t) -
h_i^1(t)\xi_i^1 \right\}\right) \; \; \; ,
\end{equation}
where $\phi(x)=0$ if $x<0$, $\phi(x)=1$ if $x>1$ and $\phi(x)=x$ otherwise,
$\gamma(t)$ is the global activity and $h_i(t)$ is the afferent current at site
$i$.
Since $E_\lambda$ is a positive semidefinite quadratic form, every local
minimum in $[0,1]^N$ is also an absolute minimum in the same interval and hence
the gradient flow converges to the ground--state of $E_{\lambda}$.
In Fig. 2 we compare the ground state energy
$F_{\lambda}$ computed analytically with that given by simulations of
Eq. (\ref{simpleupdating}) for a network of $N=1000$ neurons. It shows a
remarkable agreement between the analytic solution and the numerical
simulations.

{\bf (iii)} When $F_{\lambda}=0$ (which implies $\gamma=K$) the minimum of the
cost function defined on $[0,1]^N$ is also an absolute minimum of the same
function defined on $R^N$ and therefore its gradient vanishes. This leads  to a
very simple relation  between the activity and the stability $\Delta_i\equiv
h_i \xi_i$  at each site [upon inserting in (\ref{simpleupdating}) $\gamma=K$]
\begin{equation}
\label{actstab}
K-\Delta_i=\alpha\epsilon_i \; \; \; ,
\end{equation}
which also coincides with the expression of the afferent currents, Eq.
(\ref{fieldover}), in which $m_{\mu}=0$ for $\mu>1$.
The above relation yields straightforwardly the probability distribution of the
$\Delta$'s ${\cal P}(\Delta_i=\Delta)= {\cal P}[\epsilon_i=(K-\Delta)/\alpha]$,
which implies that, for $\alpha\leq\alpha_0(K)$, this distribution is bounded
between $\Delta=K$ and $\Delta=K-\alpha$.
For $\alpha>\alpha_0(K)$, due to the nonlinear constraint on the bounds of
the activities, the flow (\ref{gradE}) (\ref{simpleupdating}) reaches a fixed
point which does not coincide with the minimum of $E_{\lambda}$ in $R^N$,
and therefore the stabilities distribution is no longer given by
Eq. (\ref{actstab}).

{\bf (iv)} Under the initial assumption on the neuronal transfer function
[$x f(x) \geq 0$], the condition for a ground-state configuration correlated
with the presented pattern to be a fixed point of the dynamics is to
have positive $\Delta$'s at all sites. This indeed happens if  the storage
level $\alpha$ satisfies $\alpha<\alpha_{\star}(K)$ where $\alpha_{\star}(K)$
is the critical line identified by $\alpha_{\star}(K)=\min(K,\alpha_0(K))$ and
shown in Fig. 1.

It follows from Eq. (\ref{actstab}) that, when $\alpha<\alpha_{\star}(K)$,
the ground state activities are fixed points of the network dynamics
with the (nonmonotonic) transfer function $f$
\begin{equation}
\label{tf}
f(h)=\left\{
	\begin{array}{ll}
		\mbox{sgn}(h) & \parbox[t]{5cm}{if $|h|\in[0,K-\alpha]$} \\
		\mbox{sgn}(h)(K-|h|)/\alpha & \parbox[t]{5cm}
				{if $|h|\in[K-\alpha,K]$} \\
		0 & \parbox[t]{5cm}{if $|h|>K$}
	\end{array}\right.
\end{equation}
shown in Fig. 3.
In the same figure we also report, for $\alpha\!<\!\alpha_{\star}(K)$, the
equilibrium distribution of the local currents obtained by numerical
simulations performed on a network with such a dynamics. All the $\Delta$'s
belong to the interval $[K-\alpha,K]$, as expected.
It is worth noticing that, at equilibrium, only the region ${\cal R}
=[-K,-(K-\alpha)] \cup [K-\alpha,K]$ plays a role; as far as the equilibrium
properties are concerned, outside this interval the form of the transfer
function is arbitrary.
In ${\cal R}$, the dynamical behavior induced by $f$
corresponds to a regulation of the output activity at
each site. The latter turns out to be proportional to the difference between
the afferent current and a reference feedback signal equal to the global
average activity.

If the fixed points are stable, it means that for $\alpha<\alpha_{\star}(K)$
a network with the discussed dynamics is capable of stabilizing a configuration
with activity $K$ highly correlated with the retrieved pattern: the optimal
activity is $K_{opt}=0.41$ for which we have $\alpha_{\star}(K_{opt})=0.41$.
The stability of the fixed points is difficult to prove, since the distribution
of the local currents has peaks at points where the derivative of the transfer
function is discontinuous. We have checked numerically their stability for
$\alpha<\alpha_{\star}(K)$: in order to correctly initialize the system, we
have used Eq. (\ref{simpleupdating}) to find one set of initial ground-state
activities $\{\epsilon_i(0)\}$ and then took $V_i(0)=\xi_i^1 \epsilon_i(0)$.

Usually, the critical capacity is defined as an upper bound for the
presence of retrieval states correlated with the presented pattern. In
the case of a sigmoid transfer function, the critical capacity is
obtained with a finite $\sigma^2$ and yields $\sim 0.14$
\cite{kuhn,ags}. Here $\alpha_{\star}$ is derived as an upper bound for
the presence of retrieval states with $\sigma^2=0$. This means that
$\alpha_{\star}$ is a lower bound for the critical capacity, which is
expected to be higher in the region where the $\Delta$'s are strictly
positive at all sites, i.e., $K>\alpha_{\star}$ or $K>0.41$.
Interestingly enough, our results on the maximal storage capacity
$\alpha_{\star}(K_{opt})$ are very similar to an estimate obtained in
\cite{amari&co} by a  completely different method on a particular
nonmonotonic transfer function.

The question of the size of basins of attraction in such a network
remains open\cite{dyndil}.  Obviously, the condition of local stability
does not ensure that starting from an initial configuration highly
correlated with a stored memory [as, for instance, $\{
V_i(t=0)=\xi_i^\mu \}, i=1,\dots,N$], the network will converge to a
ground-state configurations belonging to the same memory.  Preliminary
numerical simulations show that the basins of attraction can be
considerably enlarged if one uses a dynamical threshold $\theta(t)$
instead of $K$ in Eq. (\ref{tf}), determined at time $t$ by the
instantaneous global activity of the network [$\theta(t)=\gamma(t)$],
given by Eq. (\ref{defact}) at time $t$. Such dynamical nonmonotonic
behavior might be seen as an effect of a regulatory mechanism of the
global activity in the network, which in real cortical networks is
supposed to be due to inhibitory interneurons.


\begin{thebibliography}{99}
\bibitem{dyndil} work on an extremely diluted network is in progress.
\end{thebibliography}
\end{document}